\documentclass[fleqn,12pt,twoside]{article}
\usepackage{espcrc1}
\usepackage{graphicx}
\usepackage[figuresright]{rotating}

\newcommand{\AmS}{{\protect\the\textfont2
  A\kern-.1667em\lower.5ex\hbox{M}\kern-.125emS}}
\begin{document}

\title{Pentaquarks in the $\overline{10}_F$- and ${10_F}$-plets}

\author{V. Dmitra\v sinovi\' c
\address{Vin\v ca Institute, lab 010, \\
        P. O. Box 522, 11001 Belgrade, Serbia \& Montenegro}
        \thanks{e-mail : dmitra@vin.bg.ac.yu}
and Fl. Stancu
\address[MCSD]{Universit\'{e} de Li\`ege,
        Institut de Physique B.5, \\
        Sart Tilman, B-4000 Li\`ege 1, Belgium} \thanks{%
e-mail : fstancu@ulg.ac.be}}
\maketitle

\begin{abstract}
We discuss the mass splittings of the pentaquark $10_F$-, and ${\overline
{10}}_F$-plet due to the quark mass differences, as well as %We discuss
the $10_F$-plet's median mass relative to that of the ${\overline {10}}_F$%
-plet of pentaquarks for both the flavour-spin and the colour-spin hyperfine
interactions, and for both parities of the pentaquark ground states. We show
that the colour-spin interaction leads to degenerate
$10_F$- and ${\overline {10}}_F$-plets when the parity is even, and a median mass splitting of
75 MeV for odd parity. The flavour-spin interaction leads to
$10_{F}-{\overline {10}}_F$ median mass splittings of 200 MeV and
40 MeV, for even and odd parities, respectively. We display
mass relations between $10_F$- and ${\overline {10}}_F$-plets
and analyze the presently known baryon resonances in this light.
\end{abstract}

\bigskip % typeset front matter

\vspace{1cm}

%\section\section{\it Intro}
\textit{Introduction} The recent wave of experimental activity led to the
observation of two purely exotic pentaquark states %resonances
\cite{NAKANO,ALT}. These are the strangeness $S=1$, $\Theta ^{+}(1540)$ and
the strangeness $S=-2$, $\Xi ^{--}(1862)$ resonances, with very small
widths, of one MeV or more, but smaller than 25 MeV and 18 MeV respectively
(limits imposed by the experimental resolution of \cite{NAKANO}and \cite{ALT}%
). Their flavour quantum numbers present indubitable evidence that they
consist of, at the very least, four valence quarks and one valence
antiquark, %may be interpreted as
i.e., that they are ``pentaquarks'' \footnote{%
Of course, this state may contain higher Fock space components, such as
septaquarks, with identical quantum numbers. We apply Occam's razor to such
components.}. The former state has been independently confirmed \cite
{BARMIN,STEPANYAN,BARTH,ASRATYAN,CLAS,HERMES,ALEEV}, whereas the latter
awaits confirmation, see Ref.  \cite{Fischer}.
At present their spins and parities are experimentally unknown, but
the chiral
soliton model ($\chi SM$), which so accurately predicted the $\Theta
^{+}(1540)$ mass and width, demands spin-parity $J^{P}={\frac{1}{2}}%
^{+}$, both for $\Theta ^{+}$ and $\Xi ^{--}$, as %being
members of the same flavour antidecuplet \cite{Diak}. Here we shall adopt
the same point of view inasmuch as both standard versions of the constituent
quark model, \textit{viz.} the flavour-spin hyperfine interaction (HFI)
%based on the pseudoscalar meson (Goldstone boson) exchange model
and the colour-spin hyperfine interaction model,
%based on the one-gluon exchange model, both
predict even parity pentaquarks as their ground states, as shown below.
For completeness, however, we shall study both parity cases and shall
discuss the relative positions of the corresponding states.

\begin{figure}[tbp]
\centerline{\includegraphics[width=4.0in,height=3.0in]
{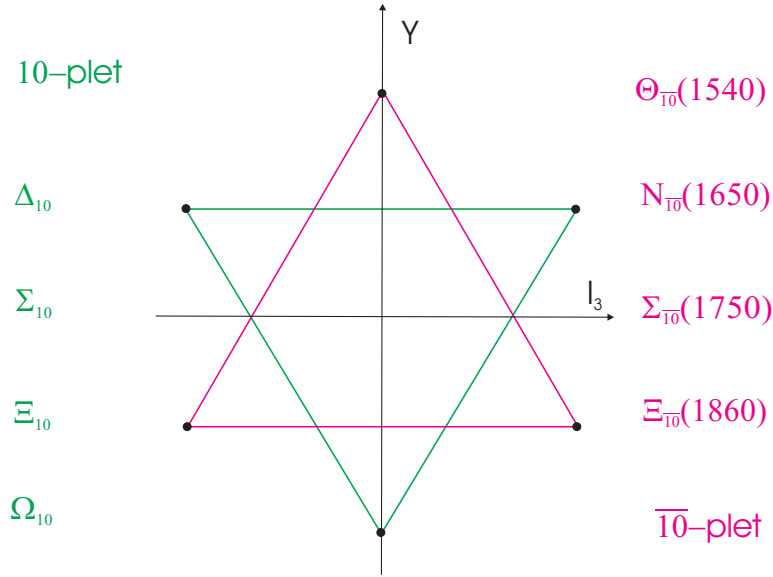}}
\caption{The weight diagrams of the $\overline {10}_F$- and the
$10_F$-plets in SU(3)$_F$ in the absence of any hyperfine interactions.}
\label{f:decu}
\end{figure}

Pentaquarks can be thought of as being constructed from a $q^{3}$
(baryon)
and a $q{\overline{q}}$ (meson) subsystem, which means that they can
be either in the $8_{F}\times 8_{F}$, or in the $10_{F}\times 8_{F}$
direct products of SU(3).
If the SAPHIR Collaboration results \cite{BARTH} can be
taken as conclusive, then the $\Theta ^{+}$ has no isospin partners and thus
(making the most conservative assumption) corresponds to top corner of the
weight diagram,  see Fig. \ref{f:decu}, of the SU$_{F}$(3)
$\overline{10}$-plet \cite{Diak}.
% \footnote{It is, of course,
%conceivable that the $\Theta ^{+}$ corresponds to top corner of the
%weight diagram of an SU$_{F}$(3) $\overline{28}$-plet, or some other
%higher dimensional multiplet, albeit that would not be a pentaquark,
%but a rather more exotic state.}.
Note that only the Clebsch-Gordan series of
$8_{F}\times 8_{F}=27_{F}+10_{F}+{\overline{10}}_{F}+2(8_{F})+1_{F}$
contains
the $\overline{10}$-plet while in the direct product decomposition
$10_{F}\times 8_{F}=35_{F}+27_{F}+10_{F}+8_{F}$ the
${\overline{10}}_{F}$
does not appear.
For this reason, for the time being, we concentrate only on
the $8_{F}\times 8_{F}$ product. Moreover, the $\Xi ^{--}(1862)$,
if its
detection is confirmed, is likely to be a member of an isoquartet,
thus
providing one (the left-hand side) bottom corner in the weight diagram of
the SU$_{F}$(3) $\overline{10}$-plet.

An immediate challenge is to give a quark model interpretation (quark
wave function) of the solitonic $\overline{10}$-plet states, as the
relation between the latter and the quark model is tenuous at best
\footnote{Indeed, in one version of the $\chi SM$, the Skyrme model,
there are no
quarks at all.}, and a dynamical explanation of the
$\overline{10}$-plet's
low mass, as well as that of the absence %/heavier mass
of other flavour multiplets. The problem lies in the large number of
possible pentaquark $\overline{10}$-plet states in the quark model,
among
which one has to find the lowest lying one, as well as in the
non-uniqueness
of the constituent quark interactions.

In the constituent quark model one usually assumes two parts to the
quark-quark interaction: (1) a long range spin-independent part that
confines quarks of any spin or flavour alike: with this interaction
all pentaquarks are degenerate and (2) a short range spin-dependent
hyperfine part that determines the mass splittings between various
spin/flavour multiplets, of which there are two ``standard'' models,
as
mentioned above. It is this second part of the interaction that ought
to
lower the even parity ${\overline{10}}_{F}$-plet's mass and keep
other
states' masses higher.

We wish to study exotic pentaquarks other than those belonging to
${\overline{10}}_{F}$ in the context of the constituent quark model(s).
The $27_{F}$-plet and the $35_{F}$-plet have the largest numbers of
real, or ortho-exotics members. Both multiplets seem to have been
eliminated as
candidates for the ``home'' of
$\Theta ^{+}$ by the SAPHIR Collaboration
results \cite{BARTH}, however.
\footnote{This does not mean that the $27_{F}$-plet and the
$35_{F}$-plet do not exist
at some higher mass, (see e. g. Ref. \cite{BGS}), which fact would
make them more difficult to detect experimentally.}

As the $1_{F}$ and the $8_{F}$s are all cryptoexotics,
they may mix,
or be confused with lower ($q^{3}$) Fock components with
identical quantum
numbers, leaving us with only $10_{F}$.
% and ${\overline {10}}_F$ for exotic pentaquarks.
\footnote{
Strictly speaking even the ${10}_{F}$-plet is not a real, or
ortho-exotic:
the $10_{F}$-plet certainly shows up in the $q^{3}$ spectra, but,
for even
parity usually with spin 3/2, in the SU(6) $(56)$-plet.
It is only in the $%
2\hbar \omega $ shell that spin 1/2, even parity $q^{3}$,
$10_{F}$-plets
appear, in the SU(6) $(56,2^{+})$ sector %$70$-plet,
but those states are sufficiently heavy, see e. g. \cite{SS},
so as not to be confused, or mixed with genuine pentaquarks.}
In particular we shall
concentrate on the $J^{P}=1/2^{+}$ pentaquark $10_{F}$-plet,
which does not
have a counterpart in the soliton models, where the decuplet
can be only in
the $J=3/2$ state, due to the Wess-Zumino term.
It is therefore reasonable
to ask first where the para-exotic \footnote{Those states
that are not pure exotics, but do not mix with octet members.}
%$1_{F}$ and the $8_{F}$s.}
members, of the $10_{F}$-plet lie in comparison with the
${\overline{10}}_{F}$-plet in the
constituent quark models.

In the absence of a strong hyperfine interaction and of SU(3)$_{F}$
symmetry breaking %these two, as well as
all of the SU(3)$_{F}$ multiplets are degenerate (in this case we have an
exact SU(6)$_{FS}$ symmetry). After ``turning on'' the hyperfine
interactions, various SU(3)$_{F}$ multiplets' %change their
energies are shifted by different amounts, but the members of individual
multiplets remain degenerate. Once the SU(3)$_{F}$ symmetry is broken,
members of various SU(3)$_{F}$ multiplets start mixing; clearly this mixing
depends on both the SU(3)$_{F}$ symmetry breaking and the (HFI induced)
energy splitting of unbroken multiplets, i.e., on the hyperfine interaction
(``SU(6)$_{FS}$ symmetry breaking'').

In order to simplify the subsequent discussion we shall do our analysis in
two steps: firstly we shall separately look at the effects of SU(3)$_{F}$
symmetry
breaking on the SU(3)$_{F}$ multiplets
in the ``free" quark model
and at those of the hyperfine interactions;
secondly we shall look at their combined effect.
Note that this is not the whole story, as we shall
not take into account the
effects of SU(3) symmetry breaking upon the HFI itself,
which need not be negligible.
This, however, is within the spirit of our ``schematic model":
Such ``higher order" corrections will have to be analysed elsewhere.
%in an ``ideal" fashion if there
%are no flavour-dependent interactions, and non-ideally otherwise.

To answer these questions we must first find the SU(3)$_{F}$ symmetry
breaking pattern for the pentaquarks in the absence of HFI, and check
if it would be possible to fit $\Theta ^{+}$(1540) and
$\Xi ^{--}$(1862) into the ${\overline{10}}_{F}$-plet.
Indeed, several authors \cite{Pakvasa} have already expressed
reservations %doubts
with regard to this option. %possibility.
After calculating the pentaquark $10_{F}$-plet and
%lie in comparison with the
${\overline{10}}_{F}$-plet spectrum in the presence of
SU(3)$_{F}$ symmetry breaking, %pattern for  %doing this,
we shall turn on the HFI and see how that affects
the whole picture.

\textit{Mass formulas for pentaquarks in the absence of HFI}
%\section{\it Mass formulas}

The SU(3) weight diagrams of the $\overline {10}_F$ and the $10_F$
are depicted in Fig. \ref{f:decu}
Each ``corner" of the $10_F$-plet describes a para-exotic pentaquark state, as
in the case of $\overline {10}_F$. We denote these pentaquarks by $%
\Delta^{++}_{10}(uuu q \overline q)$, $\Delta^{-}_{10}(ddd q \overline q)$
and $\Omega^{-}_{10}(sss q \overline q)$ respectively, whereby the quark
labels are to be understood merely as the pentaquarks' (schematic) quark
contents and not as their flavour wave functions.

Due to the large multiplicities of the $8_{F}$-plet and
% and the
%non-exotic members \footnote{Those
%members of the decimet that do mix with the octets.}
of the $10_{F}$-plets in the
pentaquark Clebsch-Gordan series,
$3_{F} \times 3_{F} \times 3_{F} \times 3_{F} \times
{\overline{3}}_{F} = 35_{F} + 3(27_{F}) + 4(10_{F}) +
2({\overline{10}}_{F}) + 8(8_{F}) + 3(1_{F})$,
(8-, and 4-fold, respectively),
% however,
%Quantum mechanics allows for complicated mixing of these states
%among themselves and with the corresponding $q^3$ states.
disentangling the mixings of these two kinds of states %multiplets
seems like a hopeless task at the present moment.%Therefore w
We rather concentrate on those
members of the decimet that do not mix with the octets.
%(we shall call them para-exotics).
%For this reason in the following we shall consider only the
%para-exotic members of the $10_{F}$-plet.
This does not mean that we may forget about the mixings of the
para-exotic decimet states among themselves or with states from
``higher" multiplets:
there are states in the $35_{F}$-plet that do mix with (at least
some of) them.
%in comparison with the ${\overline {10}}_F$-plet in the constituent
%quark models?
%Note that, since there are four distinct $10_{F}$-plets,
%there can be mixing, at least in principle, among these four
%and other states with identical flavour quantum numbers.
Note, however,
that not all of these multiplets need appear among the ground states
of any given spin and parity, e.g. there is only one octet and no
singlets among the %even-parity
$J^{P} = 1/2^{+}$ ground state $q^3$ baryons. By the same
token, a smaller number (than four and two respectively)
% indicated above)
of $10_{F}$-plets and ${\overline{10}}_{F}$-plets may actually
appear among pentaquarks, depending on the parity of the ground
state. The precise number can only be determined after complete
anti-symmetrization of the wavefunction, i.e., after full
SU(6)$_{FS}$ analysis.
%, where this number depends on the parity of the state.
%again as in the $q^3$ case: in the even
% parity case there is one, and for odd parities two octets.

The pentaquark masses must agree with the linear mass formula
$M = M_0 + c Y$, to the lowest order in the SU(3)$_F$ symmetry
breaking,
but the $M_0$ and the $c$ need not be the same for $10_F$ and
$\overline {10}_F$, or even for different $10_F$-plets.
Indeed, the simplest (``free") quark model %calculation
leads to
$M_{0}^{\mathrm{free}}(\overline {10}) = \frac{5 }{3} (m_{u} + m_{d} +
m_{s}) \simeq$ 1825 MeV and
$c(\overline {10}) = - {\frac{1 }{3}}(m_{s} - m_{d})
\simeq$ - 50 MeV.
% for both sets of ${\overline {10}}_F$-plet SU(3)
%``wave functions".
Here we used the standard strange and
up/down constituent quark masses of 465 MeV and 315 MeV,
respectively. These results are to be compared with the
``experimental" values of
$M_{0}(\overline {10}) = \frac{1 }{3} (M_{\Theta} + 2 M_{\Xi})
\simeq$ 1755 MeV and
$c(\overline {10}) = {\frac{1 }{3}}(M_{\Theta} - M_{\Xi}) \simeq$
- 107 MeV, respectively, which
are based on the measured masses of $\Theta_{\overline {10}}^+$ and
$\Xi_{\overline {10}}^{--}$. The substantial discrepancy in the
value of $c(\overline {10})$ %might perhaps
should be accounted for by the addition of a hyperfine
interaction.
%may lead to real problems, since quark
%hyperfine interactions are sufficiently strong to change the
%pentaquark's absolute mass by this amount, but not the
%value of $c(\overline {10})$, see below.

As mentioned above there are four $10_{F}$-plets in the C.G. series,
which means e. g. that in the $q^4$ subsystem there are three
(flavour) basis vectors of permutation symmetry $[31]_F$,
\footnote{Here
$[mn]$ is a partition of 4, or equivalently a Young diagram with
four boxes altogether, $m$ of which are in the first row and $n$
in the second one}, and one of symmetry $[4]_F$ (see Appendix
\ref{app:WF}).
% and one
%quasi-decimet within the $35_{F}$-plet.% in the above CG series.
%Of the four decimets, three are degenerate at %with masses
Each of the three basis vectors of symmetry $[31]_F$ give
$M(\Omega) = 2 {\bar m} + 3 m_{s}$ and
$M(\Delta) = 4 {\bar m} + m_{s}$, where
$2 {\bar m} = m_{u} + m_{d}$.
%and do not mix among themselves
%or with the $35_{F}$-plet.
This yields
$M_{0}^{\mathrm{free}}({10}) = {1 \over 3}(M_{\Omega} + 2 M_{\Delta})
= {5 \over 3}(m_{u} + m_{d} + m_{s})
\simeq$ 1825 MeV. %for the lighter ideally mixed decimet
and
$c(10) = {1 \over 3}(M_{\Delta} - M_{\Omega})
= - {2 \over 3}(m_{s} - m_{d}) \simeq$ - 100 MeV, i. e.
a twice larger splitting than predicted in the ${\overline {10}}_F$.

The fourth  $10_{F}$-plet mixes with a part of the weight diagram of
the $35_{F}$-plet which forms a decimet. After diagonalization the
mass matrix is
\begin{eqnarray}\label{OMEGA}
M_{\Omega} =
\left(\begin{array}{cc}
M_{+}\left(\Omega\right) & 0 \\
0 & M_{-}\left(\Omega\right)
\end{array}\right)~=\
%%%%%%%%%%%%%%%%%%%%%%%%%%%%%%%%%%%%%%%%%%%%%%%%%%%%%
\left(\begin{array}{cc}
5 m_{s} & 0 \\
0 & 2{\bar m} + 3 m_{s}
\end{array}\right)~.\
%%%%%%%%%%%%%%%%%%%%%%%%%%%%%%%%%%%%%%%%%%%%%%%%%%%%%%%%%%%%%%%%%%%%
\label{e:omegamassm1}
\end{eqnarray}
This shows that the mixing is ideal
in the ``free'' quark model.
Similarly for the $\Delta$ state mass matrix we get
\begin{eqnarray}\label{DELTA}
M_{\Delta} =
\left(\begin{array}{cc}
M_{+}\left(\Delta\right) & 0\\
0 & M_{-}\left(\Delta\right)
\end{array}\right)~=\
%%%%%%%%%%%%%%%%%%%%%%%%%%%%%%%%%%%%%%%%%%%%%%%%%%%%%
\left(\begin{array}{cc}
5 {\bar m} & 0 \\
0 & 4{\bar m} + m_{s}
\end{array}\right)~.\
%%%%%%%%%%%%%%%%%%%%%%%%%%%%%%%%%%%%%%%%%%%%%%%%%%%%%%%%%%%%%%%%%%%%
\label{e:deltamassm1}
\end{eqnarray}
There are two possible ways of associating these mixed states into
two decimets:

(a) one may group $M_{+}(\Delta)$ and $M_{+}(\Omega)$ into one
decimet, and $M_{-}(\Delta)$ and $M_{-}(\Omega)$ into another.
In this case we have $M_{0}({10}_{+}) =
{5 \over 3} (m_{u} + m_{d} + m_{s}) \simeq$ 1825 MeV and
$c(10_{+}) = - {5 \over 3}(m_{s} - {\bar m}) \simeq$ - 250 MeV
for the ($M_{+}(\Delta), M_{+}(\Omega)$) pair
and
$M_{0}({10}_{-}) = {5 \over 3} (m_{u} + m_{d} + m_{s}) \simeq$ 1825 MeV and
$c(10_{-}) = - {2 \over 3}(m_{s} - {\bar m}) \simeq$ - 100 MeV, for
($M_{-}(\Delta), M_{-}(\Omega)$).
One can see that $M_{0}({10}_{+})$ = $M_{0}({10}_{-})$ =
$M_{0} (\bar {10})$, but the corresponding splittings are larger
by a factor of 5 and 2 respectively than for the antidecuplet where
$c({\overline {10}})$ = - 50 MeV only.

(b) or one may group $M_{+}(\Delta)$ and $M_{-}(\Omega)$ into one
decimet, and $M_{-}(\Delta)$ and $M_{+}(\Omega)$ into another.
In this case we find two ``ideally mixed" $10_{F}$-plets,
one heavier, with a hidden $s{\bar{s}}$ pair and
$M_{0}({10}_{h}) = {1 \over 3}[4(m_{u} + m_{d}) + 7 m_{s}]
\simeq$ 1925 MeV, and
$c(10_{h}) = - {4 \over 3}(m_{s} - m_{d}) \simeq$ - 200 MeV,
and another, lighter one with a hidden light quark pair
($u{\bar{u}}$, or $d{\bar{d}}$), with
$M_{0}({10}_{l}) = 2 (m_{u} + m_{d}) + m_{s}
\simeq$ 1715 MeV and
$c(10_{l}) = - (m_{s} - m_{d}) \simeq$ - 150 MeV .
The numerical values are based on the
usual quark masses (see above).

\textit{Hyperfine interactions}
%\section{\it hyperfine interactions}
We shall work with
%two different hyperfine interactions
%(see below)
%and consider the breaking of SU(3)$_{F}$.
%; we shall show that in certain cases
%there may exist an
%analogon of the original degeneracy even in the broken SU(3)$_{F}$
%symmetry
%case and that the level (pentaquark mass) splittings are small and
%calculable.
the two most commonly used hyperfine interactions:
(1) the colour-spin (CS) model, and
(2) the flavour-spin (FS) model.
%Curiously, the lowest lying decimets in both of these models,
%for both parities, turn out to be of the non-mixing kind.
The former is based on the one-gluon exchange (OGE), and used to
be the mainstay of hadron spectroscopy
for well over 20 years; the latter is based on the pseudoscalar
(or Goldstone) boson exchange (GBE), and has been promoted over
the past eight
years or so \cite{GR96}, in response to certain phenomenological
shortcomings of the CS model in the baryon spectra.
\footnote{
The FS interaction correctly reproduces the level ordering of even and
odd parity baryon states both in non-strange and strange sectors
while
the CS model does not.} One should also remember that the FS model led to
the prediction of stable even parity heavy-flavoured pentaquarks
\cite{FS}, presently called $\Theta _{c}$ and $\Theta _{b}$,
long before the recent experimental discoveries of light-flavour
pentaquarks. \footnote{The first sighting of the charmed pentaquark
$\Theta _{c}$ has recently been reported \cite{H1}.} Even more
recently it was shown that the FS model can also accommodate the light pentaquark $%
\Theta ^{+}$ with the (minimal) quark content $uudd{\overline{s}}$
\cite{SR}
as well as the whole even parity antidecuplet \cite{CARONE,FS1}.

%At first sight it may seem that o
One must merely augment the
``free'' quark model predictions by the hyperfine-induced $10_{F}$,
and $\overline{10}_{F}$-plet mass shifts $E_{\mathrm{HF}}({10}_{F}),
E_{\mathrm{HF}}({\overline{10}}_{F})$, as follows
\begin{eqnarray}
M_{0}({10}_{F}) &=&
M_{0}^{\mathrm{free}}({10}_{F}) + E_{\mathrm{HF}}({10}_{F})
\label{e:def10} \\
M_{0}({\overline{10}}_{F}) &=&
M_{0}^{\mathrm{free}}({\overline{10}}_{F}) +
E_{\mathrm{HF}}({\overline{10}}_{F}),\
\label{e:def10bar}
\end{eqnarray}
in order to obtain the observable $10_{F}$-
and $\overline{10}_{F}$-plet median masses
$M_{0}({10}_{F}), M_{0}({\overline{10}}_{F})$.
Here $E_{\mathrm{HF}}({10}_{F}),E_{\mathrm{HF}}({\overline{10}}_{F})$
are the model-dependent hyperfine-induced mass shifts,
such as those in Eqs. (\ref{e:CSdiff}) and (\ref{e:FSdiff}) below.
This is true for states that do not mix.
But for states that do mix this is not necessarily correct,
as the HFI may change the energies of individual multiplets
depending on their flavour content and permutation symmetry.
That will not be necessary here for, as we shall see below, the
lowest states are always those which do not mix. %, so there is no
%point in discussing the mixing kind here.
The reason that the lowest lying states do not mix is that they
all originate from $q^4$ subsystems with $[31]_F$ symmetry.

\textit{The Colour-Spin model} %\section{\it CS HFI}
The schematic \footnote{Due to our neglect of spatial dependence
in the interaction Hamiltonian.}
hyperfine interaction in the CS model is the colour-spin operator
\begin{equation}  \label{CM}
{V}_{cm }\ =
 -\ {C}_{cm}\ \sum\limits_{ i\ <\ j}^{5} {\lambda }_{i}^{C}
\cdot {\lambda }_{j}^{ C} \ {\vec{\sigma }}_{i}
\cdot {\vec{\sigma }}_{j}.
\end{equation}
where the sum runs over all pairs, the particle 5 being an antiquark, i.e., $%
{\lambda}_{5}^{} \equiv - {\lambda}_{5}^{*}$. Here $\lambda _{i}^{C}$ are
the Gell-Mann matrices for colour SU(3)$_C$, and $\vec{\sigma}_i$ are the
Pauli spin matrices. From the fit to the $\Delta - N$ mass splitting $\Delta
- N = 16 {C}_{cm} \simeq $ 300 MeV one finds ${C}_{cm} \simeq 18.75$ MeV.
\footnote{%
The hyperfine Hamiltonian Eq. (\ref{CM}) expectation values in the $q^4 {%
\overline q}$ system can be calculated by way of a formula in \cite
{JAFFE,HOGAASEN}.} The results are exhibited in Table \ref{OGE} for the four
different flavour and parity cases, as derived in Appendix \ref{app:CS},
where the technical details are relegated to.

%%%%%%%%%%%%%%%%%%%%%%%%%%%%%%%%%%%%%%%%%%%%%%%%%%%%%%%%%%%%%%%%%%%
\begin{table}[tbp]
\caption{The colour-spin hyperfine interaction expectation values for the
lowest-lying even and odd parity $J= 1/2$ pentaquarks. Each state
is labelled by colour-spin, colour, spin and flavour indices representing
the dimensions of SU(6)$_{CS}$, SU(3)$_C$, SU(2)$_S$ and SU(3)$_F$
representations. The $\overline q$ defined by $|{\overline 6}_{CS},{%
\overline 3}_C,2_S,\overline {3}_F \rangle $ is coupled to the $q^4$ state
given in the first column in each case.}
\label{OGE}
\begin{tabular}{cccc}
\hline
$q^4 $ state & $q^4 {\overline q}$ state & Parity & $\langle V_{cm}
\rangle/C_{cm}$ \\ \hline
&  &  &  \\
$|210_{CS},3_C,1_S,\overline {6}_F \rangle$ & $|70_{CS},1_C,2_S,\overline
{10}_F \rangle$ & + & -40 \\
$|210_{CS},3_C,1_S,15_F \rangle$ & $|70_{CS},1_C,2_S,10_F \rangle$ & + & -40
\\
$|105_{CS},3_C,1_S,\overline {6}_F \rangle$ & $|70_{CS},1_C,2_S,\overline
{10}_F \rangle$ & - & -24 \\
$|105^{\prime}_{CS},3_C,1_S,15_F \rangle$ & $|20_{CS},1_C,2_S,10_F \rangle$
& - & -20 \\ \hline
\end{tabular}
\end{table}
%%%%%%%%%%%%%%%%%%%%%%%%%%%%%%%%%%%%%%%%%%%%%%%%%%%%%%%%%%%%%%%%%%%%%%%%%%%%%%%
From the same Table \ref{OGE} one can also see that the $10_F$-plet is
either degenerate with, or is heavier than the ${\overline {10}}_F$-plet by
\begin{equation}
{E_{\mathrm{HF}}({\overline {10}}_F) - E_{\mathrm{HF}}(10_F)} = \left\{ %
\renewcommand{\arraystretch}{2}
\begin{array}{cl}
0 & \hspace{1.1cm} \mbox{+ parity} \\
- 4 C_{cm} \simeq - 75 \, \mathrm{MeV} & \hspace{1.1cm} \mbox{- parity}
\label{e:CSdiff}
\end{array}
\right.
\end{equation}
which is the statement we set out to prove.
%%%%%%%%%%%%%%%%%%%%%%%%%%%%%%%%%%%%%%%%%%%%%%%%%%%%%%%%%%%%%

\textit{The Flavour-Spin model} %\section{\it FS HFI}
In the FS model the $qq$ and $q \overline {q}$ interactions are treated
differently. The $qq$ interaction has a flavour-spin structure and in the
following we shall employ only its schematic form \cite{GR96}:
\begin{equation}  \label{CHIRAL}
{V}_{\chi }\ =\ -\ {C}_{\chi }\ \sum\limits_{ i\ <\ j}^{4} {\lambda }_{
i}^{F} \cdot {\lambda }_{ j}^{ F} \ {\vec{\sigma }}_{i} \cdot {\vec{\sigma }}%
_{j}.  \label{VCHI}
\end{equation}
where the sum runs over $qq$ pairs only. Here $\lambda _{i}^{F}$ are the
Gell-Mann matrices for flavour SU(3)$_F$.
%, and $\vec{\sigma}_i$ are the Pauli spin matrices.
The constant $C_{\chi}$ has been determined from the $\Delta$-N mass
splitting as $C_{\chi}\simeq $ 30 MeV.

The nature of the $q \overline {q}$ interaction is different, however:
%In Ref. \cite{SR} the interaction between ${\overline s}$ and the light
%pentaquarks was assumed to be due to an $\eta$ meson exchange, in agreement
%with the study of pion decay $D^*_s \rightarrow D_s \pi^0$ \cite{LR}.
%Accordingly 
this interaction was parametrized as a spin-spin one
%interaction
without flavour dependence. 
Here we shall assume that such a potential %parametrization
holds for any light $q \overline {q}$ pair, 
%acting only among $q {\bar q}$ pairs,i.e., 
and that it does not affect the $q q$ pairs.
%, i. e., that this interaction is flavour independent.
This brings about the same amount of attraction to every SU(3)$_F$
multiplet and is only a function of the spin of the state.
One can reduce our study of the $q^4 \overline {q}$ system to the
study of the $q^4$ subsystem (see Appendix \ref{app:FS}).
The results are tabulated in Table \ref{GBE}.
%%%%%%%%%%%%%%%%%%%%%%%%%%%%%%%%%%%%%%%%%%%%%%%%%%%%%%%%%%%%%%%%%
\begin{table}[tbp]
\caption{The flavour-spin hyperfine interaction expectation values for
even and odd parity $J= 1/2$ pentaquarks. Column 2 indicates the
SU(3) flavour multiplet resulting from the coupling of $|\protect\psi^{+}_i
\rangle$ and $|\protect\psi^{-}_i \rangle$ to $\overline q$, see Appendix
\ref{app:FS}.}
\label{GBE}
\begin{tabular}{ccc}
\hline
$q^4$ state & $q^4 {\overline q}$ state & $\langle V_{\chi} \rangle/C_{\chi}$
\\ \hline
&  &  \\
$|\psi^{+}_1 \rangle$ & $\overline {10}_F$ & - 28 \\
$|\psi^{+}_2 \rangle$ & $10_F $ & - 64/3 \\
$|\psi^{-}_1 \rangle$ & $\overline {10}_F$ & - 28/3 \\
$|\psi^{-}_2 \rangle$ & $10_F $ & - 8 \\ \hline
\end{tabular}
\end{table}
%%%%%%%%%%%%%%%%%%%%%%%%%%%%%%%%%%%%%%%%%%%%%%%%%%%%%%%%%%%%%%%%%%%%%%
From there one can see that the mean mass of the $10_F$-plet is heavier than
the ${\overline {10}}_F$-plet's one
\begin{equation}  \label{BASIC}
{E_{\mathrm{HF}}({\overline {10}}_F) - E_{\mathrm{HF}}(10_F)} = \left\{ %
\renewcommand{\arraystretch}{2}
\begin{array}{cl}
- \frac{20}{3} C_{\chi} \approx - 200\, \mathrm{MeV} & \hspace{1.1cm} %
\mbox{+ parity} \\
- \frac{4}{3} C_{\chi} \approx - 40\, \mathrm{MeV} & \hspace{1.1cm} \mbox{--
parity} \label{e:FSdiff}
\end{array}
\right.
\end{equation}
unless an additional spin-spin interaction acting only on $q {\bar q}$ 
pairs is introduced, as in Ref. \cite{SR}, 
in which case an exact $10_F$-${\overline {10}}_F$ degeneracy could be
recovered. \newline

\textit{Comparison with experiment and conclusions}
%\section{\it Conclusions}
Based on the above results in the CS and FS models, we can make
predictions for the masses of the decuplet relative to the antidecuplet.
We rely on the fact that $M_{0}^{\mathrm{free}}(10_{F})$ =
$M_{0}^{\mathrm{free}}({\overline{10}}_{F})$ for the lowest states
and that the splitting in the decuplet are expected to be larger
than in the  antidecuplet.
The hyperfine interaction changes $M_0^{\mathrm{free}}$ according to Eqs.
(\ref{e:def10}), (\ref{e:def10bar}),(\ref{e:CSdiff}) and (\ref{e:FSdiff}).
This means for example that for even parity states the CS model
predicts the same median mass both for the decuplet and the antidecuplet
but in the FS model the antidecuplet median mass should be much lower.
Smaller differences appear for odd parity pentaquarks median
mass in both models. Our results can be summarized in the form of
mass relations (or ``sum rules") as follows
\begin{equation}
M_{\Theta} + 2 M_{\Xi} = M_{\Omega} + 2 M_{\Delta} + 3
\left[{E_{\mathrm{HF}}({\overline {10}}_F) - E_{\mathrm{HF}}(10_F)}
\right]
\label{e:sumrule1}
\end{equation}
and
\begin{equation}
M_{\Delta} - M_{\Omega} = \alpha (M_{\Theta} - M_{\Xi}),
\label{e:sumrule2}
\end{equation}
where $\alpha$ can be $= 2,3,4,5$, depending on the decimet
in question. We recall that the lowest-lying decimet corresponds to
$\alpha = 2$. This mass formula could be substantially altered by
the SU(3)$_F$ symmetry breaking in the hyperfine interaction, which
we have not taken into account.
Now let us look into the Particle Data
Group's (PDG) tables \cite{pdg02} to see whether  or not we can find
possible candidates for the pentaquark decuplet.
All of the presently known excited $\Omega^{-}$ states, with as yet
undetermined spins and parities, lie above 2200 MeV,
(e.g. $\Omega^{-}$(2250)).
Of all the low-lying $\Xi$ hyperons, only two
($\Xi$(1690) and $\Xi$(1950)) have spins that might
(albeit need not)
be consistent with our $J^{P} = 1/2^{+}$ requirement.
Similarly, only two observed $\Sigma$ hyperons are
allowed by this spin-parity assumption,
{\it viz.} $\Sigma$(1660), $\Sigma$(2250). We may
instantly eliminate the $\Sigma$(2250) and $\Xi$(1690)
candidates, as too heavy and too light, respectively.
The only possible ``triplet" of states
($\Sigma$(1660), $\Xi$(1950) $\Omega$(2250))
satisfies the $(10)_{F}$-plet mass ``equidistance rule"
$M_{\Xi} - M_{\Sigma} = M_{\Omega} - M_{\Xi}$, but with
an SU(3) symmetry breaking mass splitting ($c_{10}$ = - 300 MeV)
that is roughly three times bigger than
the ${\overline {10}}_F$-plet one ($c_{\overline{10}}$ = -107 MeV).
But then we remember option (b) above, in which
the parameter $\alpha$ in the
sum rule Eq. (\ref{e:sumrule2}) can take on the value of three,
which is not inconsistent with the present experimental value.
It is, however, well known that there are no $J^{P} = 1/2^{+}$
$\Delta$ states around 1360 MeV.
Finally, this decimet's mean mass $M_{0}(10_{F}) =$ 1660 MeV,
together with $M_{0}(\overline{10}_{F}) =$ 1755 MeV,
disagrees even with the sign of Eq. \ref{e:FSdiff},
for both parities.

Identification of non-exotic pentaquarks is an empirical question that
will have to be settled by experiment. If the
pentaquarks turn out to be compact hadronic states, then the question
of the failure to find experimentally pentaquarks that satisfy the
predicted mass relations will have to be addressed.
At this stage one can only say that the predicted mass
relations should not have been a surprise:
Both of these hyperfine interactions are of the two-body kind,
thus being proportional to (at most) the quadratic Casimir
operator(s) of SU(3) and/or SU(6), which do not distinguish between a
representation (multiplet) and its conjugate. Consequently any multiplet is
likely to be degenerate with its own conjugate, subject, of course, to the
afore discussed caveats.

In other words, neither of the two hyperfine interactions is taking
full advantage of the postulated symmetries, i.e., %of all
of the allowed SU(3) and/or SU(6) group theoretical structures \cite{VD}.
Thus, the (two-body) CS model certainly cannot mimic the full QCD's SU(3)
and/or SU(6) algebraic structure (irrespective of any spatial or temporal
dependence), and the (two-body) FS model does not include all possible
phenomenological interactions. One remedy to these problems that obviously
suggests itself is to include three-body forces of either the CS %\cite{VD}
or the FS kind. This remains a task for the future.

\section*{Acknowledgments}

One of us [V.D.] wishes to thank Mr. Predrag Krstono\v si\' c for his help
with the figure and the staff of Fundamental Theoretical Physics Group,
Universit\'{e} de Li\`ege, for their kind hospitality.

%%%%%%%%%%%%%%%%%%%%%%%%%%%%%%%%%%%%%%%%%%%%%%%%%%%%%%%%%%%%%%%%%%%%
\appendix{}

\section{Appendix: The flavour SU(3) ``wave functions" of the decuplet}
\label{app:WF}

Here we give the expressions of the flavour wave functions used to
derive the decimet states' %isomultiplet
masses in the  "free" quark model and the
mass matrices (\ref{OMEGA}) and (\ref{DELTA}).
In writing the explicit form of the flavour part of the decuplet
wave functions, first we construct the basis vectors of the
$q^4$ subsystem. For this purpose it is convenient to use the
Young-Yamanouchi-Rutherford basis \cite{book}
which allows one to specify the permutation symmetry of the
last two particles, here 3 and 4. Then we couple the antiquark to
the $q^4$ subsystem with the help of SU(3) Clebsch-Gordan coefficients.
Below we give the explicit form for the flavour part of four
decuplet pentaquark $\Omega$.
%\vspace{2mm}.
%The four ()linearly independent) decimet wave functions are:

\noindent
1) The state $[31]_F$ is symmetric under the permutation of particles
1 and 2 (the so called transposition (12)), but antisymmetric under
the permutation of particles 3 and 4 (transposition (34))
\begin{equation}\label{state1}
| \Omega^{-}_A \rangle =
-\frac{1}{2} | ss [u,s] {\bar u} + ss [d,s] {\bar d} \rangle
\end{equation}
2) The state $[31]_F$  symmetric  under both transpositions
(12) and (34)
\begin{equation}\label{state2}
| \Omega^{-}_S \rangle =
-\frac{1}{2 \sqrt{2}} | (ss \{u,s\} - \{u,s\} ss ) {\bar u}
+ ( ss \{d,s\} -  \{d,s\} ss) {\bar d} \rangle
\end{equation}
3) The state $[31]_F$ antisymmetric under the (12),
but symmetric under the (34) transposition
\begin{equation}\label{state3}
| \Omega^{-}_3 \rangle =
\frac{1}{2} | [u,s]ss {\bar u} + [d,s]ss {\bar d} \rangle
\end{equation}
4) The totally symmetric state $[4]_F$
\begin{equation}\label{state4}
| \Omega^{-}_{4} \rangle =
-\frac{1}{2 \sqrt{6}} | (ss \{u,s\} + \{u,s\} ss ) {\bar u}
+ ( ss \{d,s\} + \{d,s\} ss) {\bar d} - 4 ssss {\bar s} \rangle . \
\end{equation}
The normal order 1234 for quarks holds everywhere. %understood.
We use the notation $[a,b] = ab - ba $ and
$\{ a,b \} = ab + ba$, %. We use
the Young-Yamanouchi phase convention
for the permutation group, and the SU(3) CG coefficients as well
as the SU(3) phase conventions of Refs. \cite{deSwart,Hecht}.

Now recall that both the $10_F$- and the $35_F$-plet appear in the CG series
15 $\times ~{\bar 3}$ = 10 + 35 or equivalently $[4] \times [11] =
[311] + [51]$. This means that there is a subset of states of the
$35_F$-plet which have the same overall quantum numbers $Y,I,I_3$ as
athose in a decuplet. %and a similar structure to those of $10_F$.
In particular a fifth $\Omega^-$ state, %under the question,
denoted here by $|\Omega^{-}_5 \rangle$ can be obtained:
% by orthogonality. This gives
\begin{equation}\label{state5}
| \Omega^{-}_{5} \rangle =
\frac{1}{\sqrt{12}} | (ss \{u,s\} + \{u,s\} ss ) {\bar u}
+ ( ss \{d,s\} + \{d,s\} ss) {\bar d} +2 ssss {\bar s} \rangle \
\end{equation}
It is precisely this state that mixes with $| \Omega^{-}_{4} \rangle$.
This mixing gives rise to the mass matrix (\ref{OMEGA}).
The flavour basis vectors for the pentaquark $\Delta^{-}$
are obtained by the replacement $s \leftrightarrow d$ and
${\bar s} \leftrightarrow {\bar d}$ above and the corresponding mixing
gives rise to Eq. (\ref{DELTA}).

There are other ways of writing the basis states, but
the advantage of writing them in the above form is that they can then
be naturally coupled to other parts of the wave functions depending
on spin, colour or space in order to properly satisfy the Pauli
principle for the subsystem of quarks.
%%%%%%%%%%%%%%%%%%%%%%%%%%%%%%%%%%%%%%%%%%%%%%%%%%%%%%%%%%%%%%%%%%%%%
%\appendix{}

\section{Appendix: The colour-spin model}
\label{app:CS}

We start by using SU(6) representations for the pentaquarks. In the OGE
model we consider the direct products SU(2)$_S$ $\times$ SU(3)$_C$, as
subgroups of SU(6). Then the $q^3$ and the $q {\overline q}$ subsystems are
described by the 56 and the 35 irreps of SU(6) and for the
$q^4 {\overline q}
$ system we get the SU(6) Clebsch-Gordan series
\begin{equation}  \label{CG}
56 \times 35 = 56 + 70 + 700 + 1134
\end{equation}
First we have to consider the compatibility of the symmetry of the flavour
part of the $q^4$ subsystem with either $10_F$- or ${\overline {10}}_F $%
-plets of the $q^4 \overline q$ system. We have the following four cases:
\newline

a)$J^P = 1/2^+$ pentaquarks belonging to the $\overline {10}_F$-plet.
\newline
The SU(3)$_F$ symmetry of the $q^4$ subsystem is $[22]_F$ in order to give
rise to the $\overline {10}_F$-plet because $[22]_F \times [11]_F =
\overline {10}_F + 8_F$, where $[f]$ represents a given Young tableau of
partition $[f]$ and O,C,F and S stand for the orbital, colour, flavour and
spin degrees of freedom, respectively. Note that from this decomposition of
the direct product it follows that $\overline {10}$ and $8$ are degenerate
if SU(3)$_F$ is an exact symmetry. The lowest state of even parity is $%
[31]_O$, containing one quark in the $p$-shell. This gives $[31]_O \times
[22]_F$ = $[31]_{OF} + [211]_{OF}$. The $[211]_{OF}$ part is combined with
its conjugate $[31]_{CS}$ in order to obtain a totally antisymmetric $q^4$
state. Here $[31]_{CS}$ has dimension 210 in SU(6) and it is obtained from
the inner product $[31]_{CS}= [211]_C \times [22]_S$ which means that the $%
q^4$ subsystem has spin zero, so that the total spin is $S = S_{\overline q}
= 1/2$. The total angular momentum is therefore $J = 1/2$ or $3/2$ and the
interaction Eq. (\ref{CM}) cannot distinguish between them. In SU(6)
notation the coupling to the antiquark described by the $\overline 6$
representation gives $\overline 6 \times 210 = 1134 + 56 + 70$, compatible
with the relation (\ref{CG}). The most favourable symmetry multiplet is $70$
as implied by the formula given in Ref. \cite{JAFFE} or\cite{HOGAASEN}. This
state has also been considered in Ref. \cite{CHEUNG} and is the lowest
even parity pentaquark state in the CS model. % where it was
%estimated by how much the diquark-diquark-quark model \cite{JW} or the
%diquark-quark model \cite{KL} underestimate the chromomagnetic attraction by
%neglecting intercluster interactions.
\newline

b)$J^P = 1/2^+$ pentaquarks belonging to the $10_F$-plet.\newline
The SU(3)$_F$ symmetry of the $q^4$ subsystem is $[31]_F$ which gives rise
to $15_F \times {\overline 3}_F= 10_F + 8_F +27_F$, all these multiplets
being degenerate if the SU(3)$_F$ symmetry is exact. In the orbital and the
flavour space the direct product is $[31]_O \times [31]_F$ = $[4]_{OF} +
[31]_{OF} + [22]_{OF} + [211]_{OF}$. From here we choose $[211]_{OF}$ which
combined with $[31]_{CS}$ gives the most favourable $q^4$ totally
antisymmetric state. This implies that the SU(6) direct product obtained
from the coupling to the antiquark is the same as for the case a) and the
most favourable multiplet is again 70. \newline

c) $J^P = 1/2^-$ pentaquarks belonging to the $\overline {10}_F$-plet.
\newline
The symmetry of the $q^4$ subsystem in the SU(3)$_F$ space is $[22]_F$, like
in the case a). The lowest state of odd parity is $[4]_O$ so one gets $%
[4]_O \times [22]_F$ = $[22]_{OF}$. In order to obtain a totally
antisymmetric wave function for $q^4$ one must combine the $[22]_{OF}$ part
with its conjugate in the CS space, the $[22]_{CS}= [211]_C \times [22]_S$
state, of dimension 105 in SU(6). Then the coupling of the antiquark gives ${%
\overline 6} \times 105 = 560 + 70$. The multiplet with the most favourable
symmetry is 70. This state appears as state V in Ref. \cite{HOGAASEN} and is
the lowest odd parity pentaquark state in the CS model.\newline

d) $J^P = 1/2^-$ pentaquarks belonging to the $10_F$-plet.\newline
The SU(3)$_F$ symmetry of the $q^4$ subsystem is $[31]_F$ as in case b). The
lowest state of odd parity is $[4]_O$ so one gets $[4]_O \times [31]_F$
= $[31]_{OF}$. Then the Pauli principle requires to combine it with $%
[211]_{CS}$. The coupling to the antiquark gives ${\overline 6} \times
105^{\prime}= 540 + 70 + 20$. The most favourable multiplet is 20. \newline

Table \ref{OGE} lists all the above $q^4 {\overline q}$ states together with
the states of the corresponding $q^4$ subsystem. Note that the SU(6)
representations associated to $q^4 {\overline q}$ and $q^4$ are, in linear
combinations, consistent with the considerations made in Ref. \cite{BS}.
This indicates that the even parity antidecuplet and decuplet states are
degenerate. The odd parity ones are not. Moreover even though the
contribution of the CS hyperfine attraction is roughly two times larger for
even parity states than for odd parity ones,
%the CS analogon of Eq. (\ref{DELFS}),
%with ${C}_{cm} \approx 18.75$ MeV, reads
the extra unit of orbital excitation $\hbar \omega \simeq $ 500 MeV \cite
{GR96}, carried by the even parity state leads to
\begin{equation}  \label{DELCS}
{E^+ - E^-} = \left\{ \renewcommand{\arraystretch}{2}
\begin{array}{cl}
\frac{1}{2} \hbar \omega - 16 C_{cm} \approx - 50~ \mathrm{MeV} & \hspace{%
1.1cm} \mbox{antidecuplet} \\
\frac{1}{2} \hbar \omega - \frac{40}{3} C_{cm} \approx~~ 0 & \hspace{1.1cm} %
\mbox{decuplet}
\end{array}
\right.
\end{equation}
with ${C}_{cm} \simeq 18.75$ MeV. This schematic estimate implies that in
the antidecuplet case the $1/2^+$ state is expected somewhat below the $1/2^-
$ state, in agreement with Ref. \ \cite{JM}, whereas in the decuplet channel
they would be practically degenerate.

\section{Appendix: The flavour-spin model}

\label{app:FS} %{\bf I. The GBE model} \\

%%%%%%%%%%%%%%%%%%%%%%%%%%%%%%%%%%%%%%%%%%%%%%%%%%%%%%%%%%%%%%%%%%%%%

The overall parity is determined by that of the $q^4$ subsystem. The
available SU(6) representations describing the $q^4$ subsystem are given by
the direct product decomposition
\begin{equation}
6 \times 6 \times 6 \times 6 = 126 + 3 (210) + 2(105) + 3(105^{\prime}) + 15
\end{equation}
In the GBE model the lowest totally antisymmetric $J^P = {1/2}^+$ states
constructed in the FS coupling scheme are given by
\begin{equation}  \label{STATE1}
%\left.{
\left|{\psi^{+}_1} %\right.}
\right\rangle =
%\left.{
\left|[{31}]_O
 {\left[{211}\right]}_{C} \left[{{1}^{4}}\right]_{OC}\ ;
[{22}]_{F} [22]_{S} [{4}]_{FS} %\right.}
\right\rangle
\end{equation}

\begin{equation}  \label{STATE2}
%\left.{
\left|{\psi^{+}_2} %\right.}
\right\rangle\ =
%\left.{
\left|
\left[{31}\right]_{O}
\left[{211}\right]_{C}
\left[{1}^{4}\right]_{OC}\ ;
\left[{31}\right]_{F} \left[{31}\right]_{S}
\left[{4}\right]_{FS} %\right.}
\right\rangle
%\right)
\end{equation}
In each case the colour part is uniquely defined. It gives rise to a
totally
antisymmetric OC state if combined with $[31]_O$ which contains one
$p$-shell quark state. Together with the parity of the antiquark this
leads to L=1 even parity states. As the FS part is totally symmetric
one obtains
totally antisymmetric $q^4$ states. These states were for the first time
considered in Ref. \cite{FS} in the context of even parity heavy
pentaquarks, presently denoted in the literature by $\Theta_c$ and $\Theta_b$%
. These were also the two states used in Ref. \ \cite{SR}. The first has
spin S = 0 and the second S = 1. The coupling to the antiquark spin leads to
a total S = 1/2 for both and the coupling to $L$ = 1 gives $J$ = 1/2 or 3/2.

The SU(6) flavour-spin state $[4]_{FS}$ is totally symmetric i. e. it
belongs to the representation (126) of SU(6). In this situation the only
possible SU(6) representations of a $q^4 {\overline q}$ system are given by
\begin{equation}
126 \times \overline 6 = 700 + 56
\end{equation}
It follows that the above states are compatible only with the (700) SU(6)
representation which contains both ${\overline {10}}_F$ and ${10}_F$ having $%
J$= 1/2. The state $| \psi^{+}_1 \rangle $ corresponds to ${\overline {10}}_F
$ and $| \psi^{+}_2 \rangle $ to ${10}_F$. As Table \ref{GBE} indicates
these states are not degenerate. An additional spin-spin interaction, such
as the one considered in Ref. \ \cite{SR}, with an adequate strength could
however make these two states degenerate.
%of the spin-spin interaction, by lowering the ${10}_F$-plet.

By analogy $J^P = {1/2}^-$ states can be constructed as
\begin{equation}  \label{STATE3}
%\left.{
\left|{\psi^{-}_1} %\right.}
\right\rangle
= \left.{\left|
\left[{4}\right]_{O}
\left[{211}\right]_{C}
\left[{211}\right]_{OC}\ ;
\left[{22}\right]_{F}
\left[{31}\right]_{S}
\left[{31}\right]_{FS} \right.}\right\rangle .
\end{equation}
and
\begin{equation}  \label{STATE4}
%\left.{
\left|{\psi^{-}_2} %\right.}
\right\rangle =  \left.{\left|
\left[{4}\right]_{O}
\left[{211}\right]_{C}
\left[{211}\right]_{OC} ;
\left[{31}\right]_{F}
\left[{22}\right]_{S}
\left[{31}\right]_{FS} \right.}\right\rangle .
\end{equation}
where ${\left[{31}\right]}_{FS}$ is the SU(6) representation (210) which
leads to
\begin{equation}
210 \times \overline 6 = 1134 + 56 + 70
\end{equation}
in the $q^4 {\overline q}$ system. The state $|\psi^{-}_1 \rangle $ is
compatible with the representation (1134), the only one which contains ${%
\overline {10}}_F$, and $|\psi^{-}_2 \rangle $ is compatible with (1134),
(56) or (70)-plet. From Table \ref{GBE} one can see that these states are
also not degenerate. These states have been considered in Ref. \ \cite{GE98}
in the context of odd parity heavy pentaquarks containing $c $ or $b$
antiquarks.

One can see that for both $10_F$- and ${\overline {10}}_F$-plets, the
even parity state lies far below the odd parity one. Taking into
account that the even parity states contain one unit of orbital
excitation $\hbar \omega \simeq $ 500 MeV \cite{GR96} and using Table \ref
{GBE} with $C_{\chi} \simeq $ 30 MeV one obtains %\newline
\begin{equation}  \label{DELFS}
{E(\psi^+) - E(\psi^-)} = \left\{ \renewcommand{\arraystretch}{2}
\begin{array}{cl}
\frac{1}{2} \hbar \omega - \frac{56}{3} C_{\chi} \approx - 310\, \mathrm{MeV}
& \hspace{1.1cm} \mbox{antidecuplet} \\
\frac{1}{2} \hbar \omega - \frac{40}{3} C_{\chi} \approx - 150\, \mathrm{MeV}
& \hspace{1.1cm} \mbox{decuplet}
\end{array}
\right.
\end{equation}
\noindent Thus, for odd parity pentaquarks, both the antidecuplet and
the decuplet are expected to be far above the threshold and highly unstable.

%\centerline{\bf Acknowledgement}

%%%%%%%%%%%%%%%%%%%%%%%%%%%%%%%%%%%%%%%%%%%%%%%%%%%%%%%%%%%%%%%%%%%%%%%%

%%%%%%%%%%%%%%%%%%%%%%%%%%%%%%%%%%%%%%%%%%%%%%%%%%%%%%%%%%%%%%%%%%%%%%%%%%%%%

\end{document}